\documentclass[10pt,conference]{IEEEtran}
\IEEEoverridecommandlockouts
\usepackage{cite}
\usepackage{amsmath,amssymb,amsfonts}
\usepackage{algorithmic}
\usepackage{graphicx}
\usepackage{textcomp}
\usepackage{xcolor}
\usepackage{glossaries}
\usepackage{siunitx}
\usepackage{subcaption}
\usepackage{textpos}
\def\BibTeX{{\rm B\kern-.05em{\sc i\kern-.025em b}\kern-.08em
    T\kern-.1667em\lower.7ex\hbox{E}\kern-.125emX}}

\newacronym{VR}{VR}{Virtual Reality}
\newacronym[plural=HMDs,longplural=Head-Mounted Displays]{HMD}{HMD}{Head-Mounted Display}
\newacronym{SoC}{SoC}{System-on-a-Chip}
\newacronym{QoE}{QoE}{Quality of Experience}
\newacronym{HR2LLC}{HR2LLC}{High-Rate and High-Reliability Low-Latency Communications}
\newacronym{RSSI}{RSSI}{Received Signal Strength Indicator}
\newacronym{MCS}{MCS}{Modulation and Coding Scheme}
\newacronym{COTS}{COTS}{Commercial Off-The-Shelf}
\newacronym{LoS}{LoS}{Line of Sight}
\newacronym{MTU}{MTU}{Maximum Transmission Unit}
\newacronym{ROS}{ROS}{Robot Operating System}
\newacronym[plural=APs,longplural=Access Points]{AP}{AP}{Access Point}

\graphicspath{{./images/}}

\makeatletter
\newcommand{\linebreakand}{%
  \end{@IEEEauthorhalign}
  \hfill\mbox{}\par
  \mbox{}\hfill\begin{@IEEEauthorhalign}
}
\makeatother

\begin{document}

\title{Opportunities and Challenges for Virtual Reality Streaming over Millimeter-Wave: An Experimental Analysis
}

\author{
    \IEEEauthorblockN{Jakob Struye\IEEEauthorrefmark{1}, Hemanth Kumar Ravuri\IEEEauthorrefmark{2}, Hany Assasa\IEEEauthorrefmark{3}, Claudio Fiandrino\IEEEauthorrefmark{4},\\Filip Lemic\IEEEauthorrefmark{5}, Joerg Widmer\IEEEauthorrefmark{4}, Jeroen Famaey\IEEEauthorrefmark{1}, Maria Torres Vega\IEEEauthorrefmark{2}}\\
    \IEEEauthorblockA{\IEEEauthorrefmark{1}University of Antwerp - imec, Antwerp, Belgium. Email: \{firstname\}.\{lastname\}@uantwerpen.be }
    \IEEEauthorblockA{\IEEEauthorrefmark{2}Ghent University - imec, Ghent, Belgium. Email: \{firstname\}.\{lastname\}@ugent.be}
    \IEEEauthorblockA{\IEEEauthorrefmark{3}Pharrowtech, Leuven, Belgium. Email: \{firstname\}@pharrowtech.com}
    \IEEEauthorblockA{\IEEEauthorrefmark{4}IMDEA Networks Institute, Leganes, Spain. Email: \{firstname\}.\{lastname\}@imdea.org}
    \IEEEauthorblockA{\IEEEauthorrefmark{5}Universitat Polit\`ecnica de Catalunya, Barcelona, Spain. Email: \{firstname\}.\{lastname\}@upc.edu}
}
\IEEEoverridecommandlockouts

\maketitle

\IEEEpubidadjcol

\begin{abstract}
Achieving extremely high-quality and truly immersive interactive Virtual Reality (VR) is expected to require a wireless link to the cloud, providing multi-gigabit throughput and extremely low latency. A prime candidate for fulfilling these requirements is millimeter-wave (mmWave) communications, operating in the \SI{30}{} to \SI{300}{\giga\hertz} bands, rather than the traditional sub-\SI{6}{\giga\hertz}. Evaluations with first-generation mmWave Wi-Fi hardware, based on the IEEE 802.11ad standard, have so far largely remained limited to lower-layer metrics. In this work, we present the first experimental analysis of the capabilities of mmWave for streaming VR content, using a novel testbed capable of repeatably creating blockage through mobility. Using this testbed, we show that (a) motion may briefly interrupt transmission, (b) a broken line of sight may degrade throughput unpredictably, and (c) TCP-based streaming frameworks need careful tuning to behave well over mmWave.
\end{abstract}

\begin{textblock}{180}(0,6)
    \begin{tiny}
    \copyright\ 2022 IEEE. Personal use of this material is permitted. Permission from IEEE must be obtained for all other uses,\\
    \vspace{-6mm}\\
    in any current or future media, including reprinting/republishing this material for advertising or promotional purposes,\\
    \vspace{-6mm}\\
    creating new collective works, for resale or redistribution to servers or lists, or reuse of any copyrighted component\\
    \vspace{-6mm}\\
    of this work in other works.
    \end{tiny}
    \end{textblock}
\section{Introduction}
\begin{figure*}[t]
    \centering
    \includegraphics[width=0.255\linewidth]{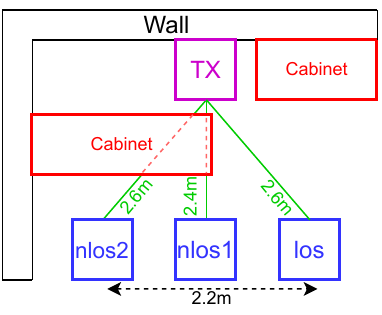}
    \includegraphics[width=0.3625\linewidth]{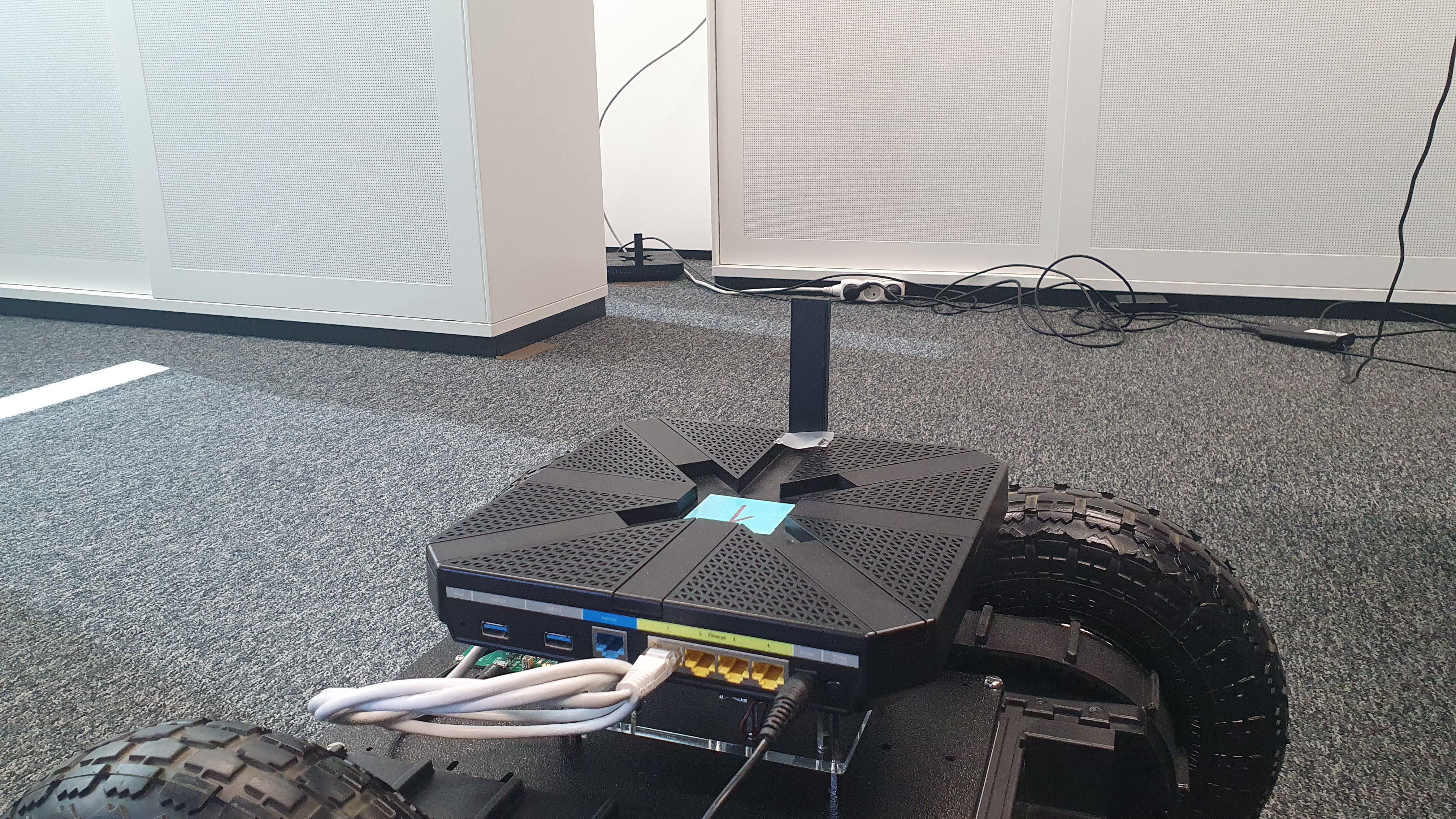}
    \includegraphics[width=0.3625\linewidth]{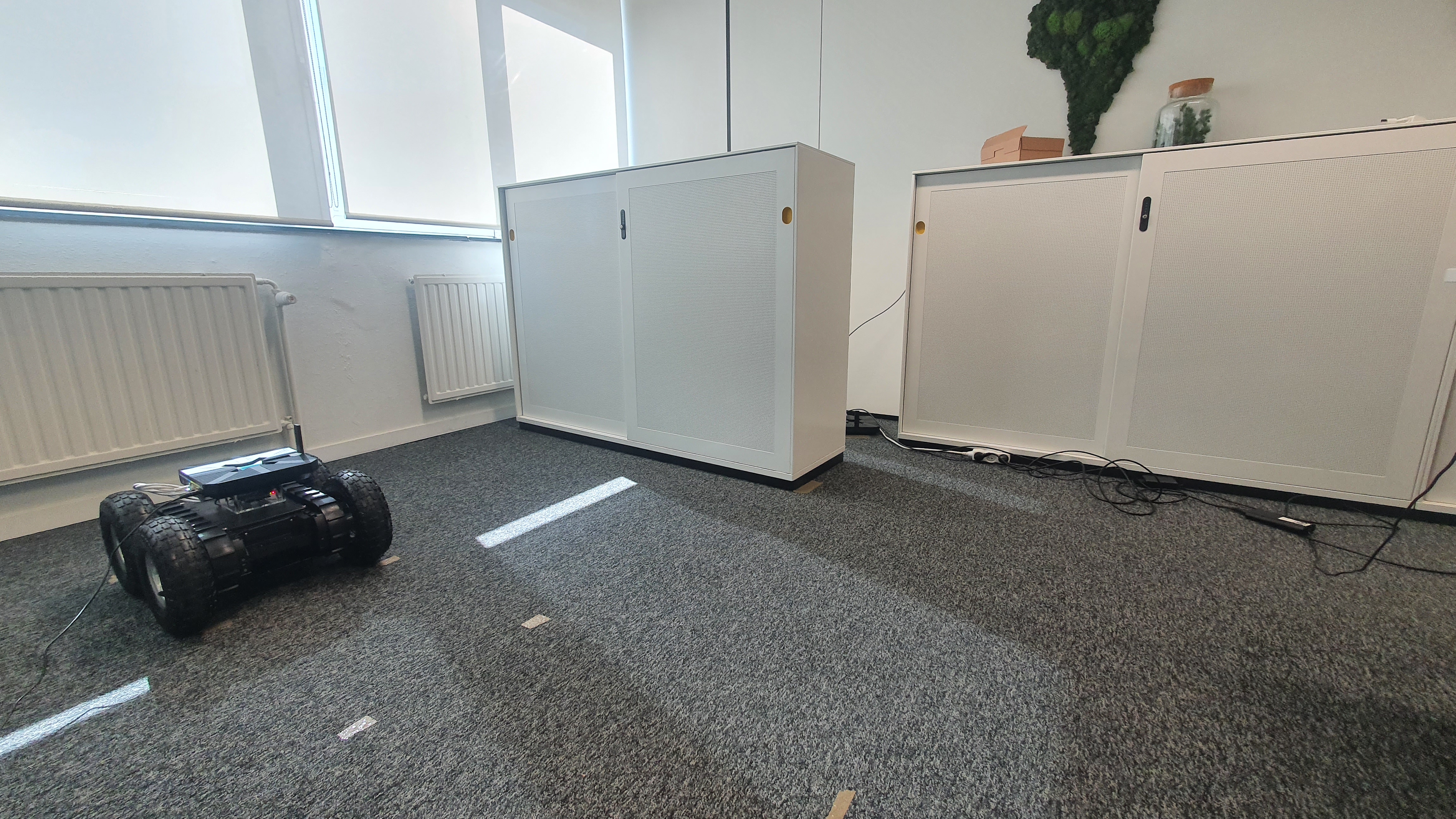}

    \caption{A top-down diagram of the testbed setup, and two pictures with the receiver in different positions.}
    \label{fig:pic}
\end{figure*}
In recent years, the market for \gls{VR} \glspl{HMD} has been shifting from tethered devices, with off-device content generation, to stand-alone devices, generating content fully on-device~\cite{VROverview}. These stand-alone devices offer increased immersiveness, as there are no cables hindering the user~\cite{VRChallengesMmWave}. However, stand-alone devices also entail a number of restrictions. Graphical fidelity is limited by the capabilities of the device. Furthermore, on-device content generation eliminates several potential interactive \gls{VR} use-cases, such as virtual meetings~\cite{VRMeetings} and remote operation~\cite{VRRemoteOp}. A wirelessly connected \gls{HMD} may be able to alleviate this: no cables can hinder the user, but content generation can still be offloaded. A future wirelessly connected \gls{HMD} offering extreme graphical quality for interactive experiences does incur strict network requirements~\cite{VRChallengesMmWave}. Specifically, such a deployment is expected to require a multi-gigabit link, capable of delivering video content at extremely low latency. When streaming content per-frame, one video-frame must be delivered fully within \SI{7}{\milli\second} to maintain an optimal \gls{QoE}~\cite{VRLatency}. Furthermore, the network must be highly reliable, as even modest packet loss is detrimental to the \gls{QoE}. Overall, these requirements are known as \gls{HR2LLC}~\cite{HR2LLC}. A prime candidate for offering this performance is Millimeter-Wave (mmWave) communications, which makes use of the \SI{30}{} to \SI{300}{\giga\hertz} bands~\cite{VRChallengesMmWave}. Given the abundance of spectrum in this range, extremely high throughput and low latency communications become feasible. However, these higher frequencies are susceptible to blockage, and the signal attenuates quickly with distance. As such, energy must be focused in specific directions, ensuring sufficiently high signal strength where desired~\cite{MmWaveBook}. This process, called \textit{beamforming}, is especially challenging in highly mobile environments, where it must adapt continuously~\cite{PIA}. Mobility and blockage are both expected to occur in \gls{VR} environments, as users often walk around and turn during these experiences. Furthermore, objects within the physical environments, or even the users' bodies can serve as obstructions causing blockage~\cite{MmWaveBlockage}.

In this work, we present an experimental assessment of the capabilities and challenges of \gls{VR} over mmWave. Some previous works have evaluated mmWave's capabilities for \gls{VR} streaming through models, simulation and experiments~\cite{VRMmWaveSim1, VRMmWaveSimExp1, VRMmWaveExpMoVR, VRMmWaveExp2, mmwaveVRNoMotion,miDroid}. Uniquely, this work employs \gls{COTS} IEEE 802.11ad routers, with which we built a novel testbed for mmWave under repeatable mobility. Using this testbed, we introduce several levels of blockage. Furthermore, as these routers also support lower-frequency legacy Wi-Fi, we can compare mmWave's unique capabilities and challenges to legacy Wi-Fi. Previous experimental evaluations of these TP-Link Talon AD7200 routers have traditionally focused on lower-layer metrics~\cite{Talon1,Talon2,Talon3,Talon4}. Some evaluations consider transport layer metrics such as TCP throughput~\cite{TalonTransport2,TalonTransport1}, but evaluations rarely focus on the application layer. In contrast, this paper analyses application-layer performance of \gls{VR} streaming over mmWave, using a modern streaming framework based on HTTP/2 over TCP. 

\section{Testbed}
Our testbed, shown in Fig.~\ref{fig:pic}, was devised to resemble a real wireless connected \gls{VR} environment. In this section, we describe the setup from a hardware and software perspective.
\subsection{Hardware}
The wireless link consists of two TP-Link Talon AD7200\footnote{https://www.tp-link.com/us/home-networking/wifi-router/ad7200/} mmWave routers, connected in a master-client network. One router is placed statically near the corner of a \SI{5.5}{}$\times$\SI{11}{\meter} room, while the other is mounted on a Rover Robotics 4WD Rover Pro\footnote{https://roverrobotics.com/products/4wd-rover-pro} robot, enabling repeatable mobility patterns for accurate evaluation of the impact of mobility. To create interference, we partially surround the static router with metal cabinets, such that the experimenter can break or restore \gls{LoS} by driving the robot around. A laptop serves as sender, while an ADLINK Vizi-AI devkit board, mounted atop the robot, performs the role of receiving \gls{HMD}. Each is connected to its router over Ethernet. The robot can be controlled over a separate wireless link using \SI{2.4}{\giga\hertz} legacy Wi-Fi, entirely disjoint from the data path under evaluation. Finally, we note that, in this iteration of the testbed, a single cable is tethered to the robot, to power the router. In a future iteration, the platform could be made fully wireless by connecting the router to the \SI{294}{Wh} battery already powering the robot and devkit.
\subsection{Software}
\begin{figure*}[t]
    \begin{minipage}[b]{.64\textwidth}
        \centering
        \includegraphics[width=0.49\linewidth]{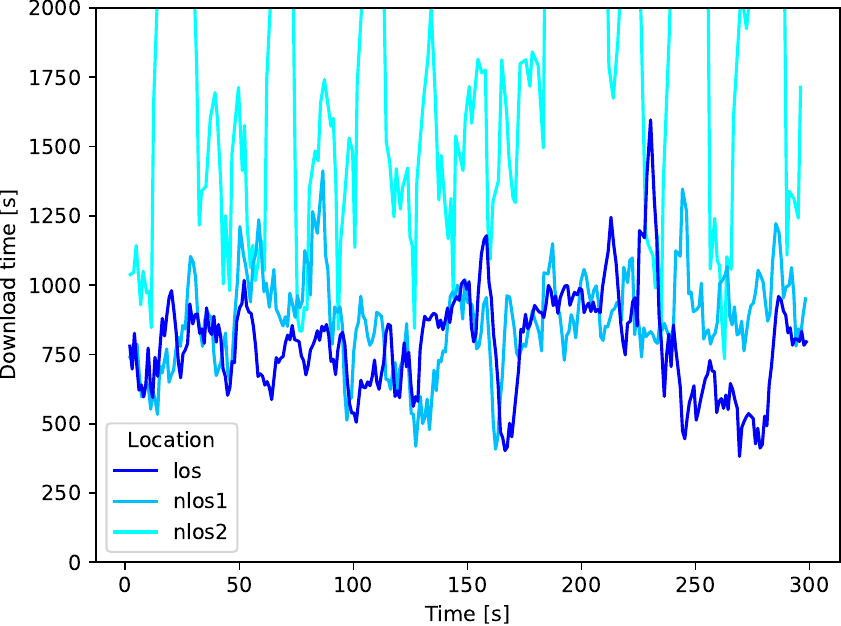}
        \includegraphics[width=0.49\linewidth]{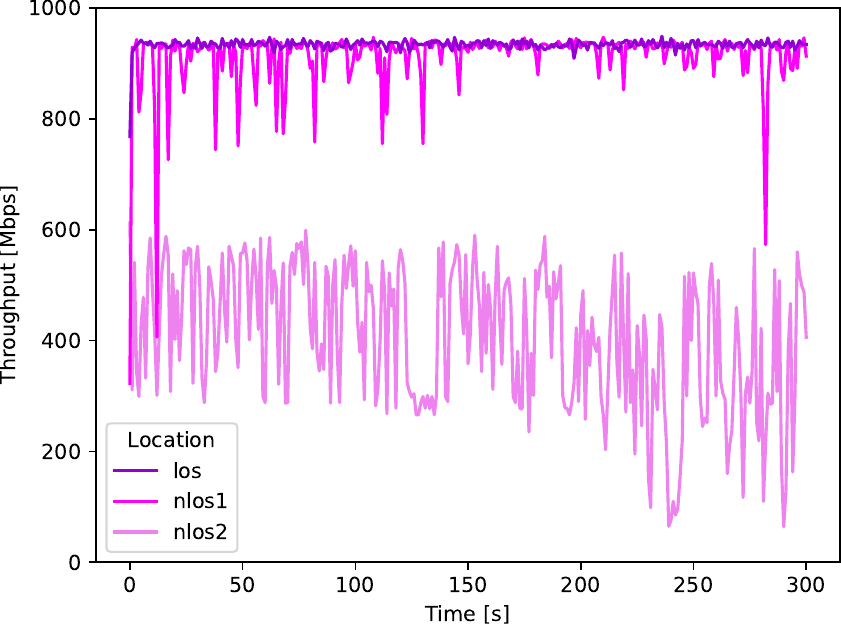}
        \caption{Streaming (left) and \texttt{iPerf} (right) performance without mobility.}
        \label{fig:base}
    \end{minipage}
    \begin{minipage}[b]{.35\textwidth}
        \includegraphics[width=\linewidth]{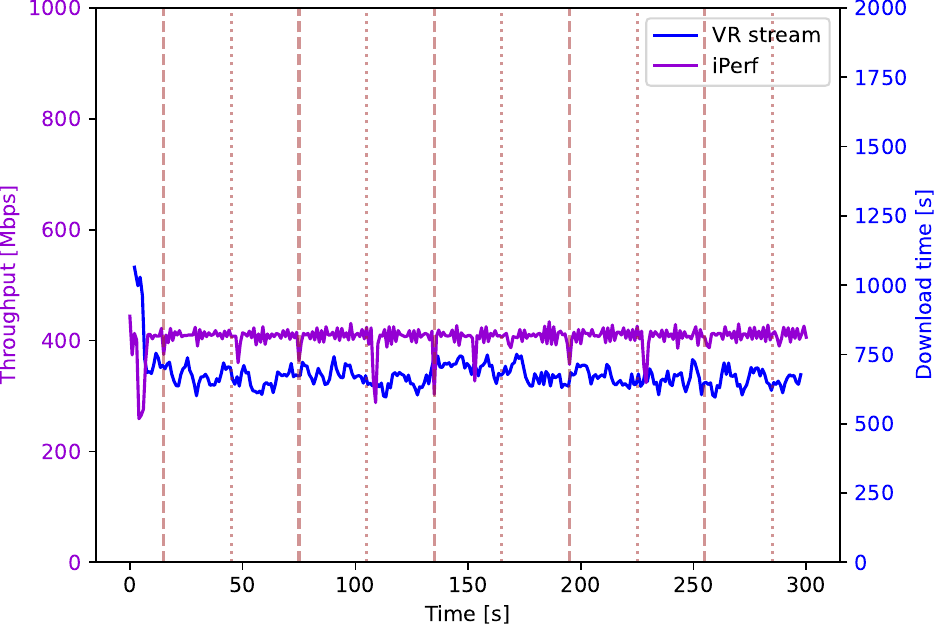}
        \caption{Performance with mobility at \SI{5}{\giga\hertz}.}
        \label{fig:legacy}
    \end{minipage}
\end{figure*}
As an operating system, both routers run mainline OpenWrt 21.02.1. In contrast, other evaluations of this router run a custom fork of this operating system, designed specifically for this router\footnote{https://seemoo.de/talon-tools/}. Independent from this fork, full support for this router was added to mainline OpenWrt recently. Crucially, only this version includes several patches for misbehaving processes, which would intermittently hog CPU resources, blocking wireless data transmission entirely for some milliseconds every few seconds\footnote{https://forum.openwrt.org/t/possible-cause-of-r7800-latency-issues/12872}. This may have significantly impacted performance in previous evaluations. In addition, we noted severe performance degradation on the routers whenever layer 3 routing was occurring. By default, OpenWrt will not allow an Ethernet link and client-mode wireless link to be in the same subnet. To avoid routing in this case, we used the \texttt{relayd} package\footnote{https://openwrt.org/docs/guide-user/network/wifi/relay\_configuration} to expose the wireless subnet over the Ethernet link, eliminating the need for routing. Experimentation showed that this method, combined with a \gls{MTU} along the entire path of \SI{1986} (the highest the mmWave chipset supports), was necessary to achieve gigabit throughput over the mmWave link. As the Ethernet interfaces are restricted to \SI{1}{Gbps}, higher throughput remains unattainable, despite the mmWave chipset supporting a theoretical PHY-level throughput of \SI{2310}{Mbps}. For beamforming, the mmWave chipset relies on a pre-defined \textit{codebook} of patterns, each focusing energy in different directions. When needed, all patterns are tested and the best-performing one is selected. Any other settings are left to their defaults: the network is set to channel 1 (\SI{2.16}{\GHz} bandwidth), and the chipset dynamically selects a TX power of at most \SI{10}{dBm}~\cite{Talon2}.

By default, the robot is controlled using joysticks connected over Bluetooth. To ensure consistent, repeatable motion, we instead employ emulated joysticks running on-device, instructing the robot to drive on a pre-defined path. 

In order to stream VR video over HTTP/2 using TCP, the sender hosts a simple web server which stores the video. It consists of eight point clouds in the form of temporal chunks called segments. To generate the volumetric media content, the 8i data set was used~\cite{d20178i}. To this end, we have taken the scene generation approach suggested in~\cite{van2019towards}. To increase the bandwidth requirements of the content, each of the four dynamic point cloud sequences was included twice to create a scene with eight point cloud objects. We have encoded the dynamic point cloud sequences individually, segmented them temporally into one-second-long segments, and made them available at the server. The average encoded bitrate of the sequence with eight objects is \SI{250}{Mbps}. Furthermore, at client side a simple headless client is implemented on the devkit board. This client uses a mechanism similar to Dynamic Adaptive Streaming over HTTP (DASH) to request the VR content~\cite{van2019towards}. As maximal \gls{QoE} requires a consistently high graphical quality, the client requests only the highest available quality. Furthermore, the adaptive quality mechanism would partially obscure the impact of blockage in this initial evaluation. As this evaluation considers pre-generated content, the client has a buffer length of one second. It requests the encoded segments of the eight point cloud objects and fills the buffer. Given the buffer length, each set of segments must arrive within one second of being requested to avoid content freezes, which would drastically impact \gls{QoE}. If a download is delayed beyond one second, the next segments are not requested until the previous download is finished, so delays cannot compound. In addition to \gls{VR} content, we also evaluate performance using a simple TCP stream with \texttt{iPerf2}.

\section{Experiments}
In this section, we present a set of experiments carefully crafted to evaluate the opportunities and challenges in using mmWave for \gls{VR} streaming, taking mobility and blockage into account. Every experiment is performed for 5 minutes, and is repeated with both \gls{VR} streaming and \texttt{iPerf} traffic.
To design the mobility patterns, we carefully selected three locations in the environment, as shown in Fig.~\ref{fig:pic}:
\begin{itemize}
\item \texttt{los}, where there is a clear \gls{LoS} between the two routers
\item \texttt{nlos1}, where the edge of the cabinet breaks the \gls{LoS}
\item \texttt{nlos2}, where the middle of the cabinet breaks the \gls{LoS}.
\end{itemize}
\begin{figure*}[t]
    \begin{minipage}[b]{.325\textwidth}
        \centering
        \begin{subfigure}[t]{\linewidth}
        \includegraphics[width=\linewidth]{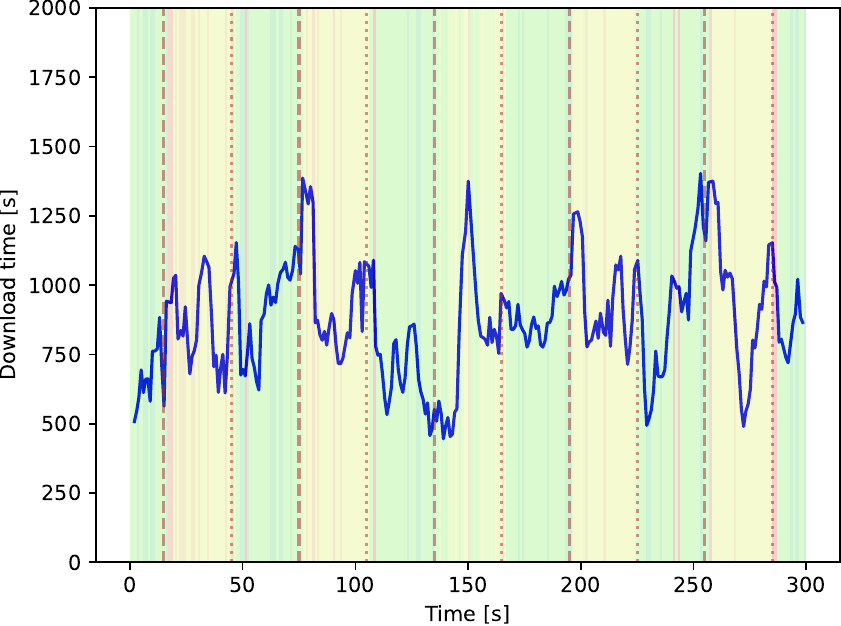}
        \caption{VR streaming}
        \end{subfigure}\\
        \begin{subfigure}[t]{\linewidth}
        \includegraphics[width=\linewidth]{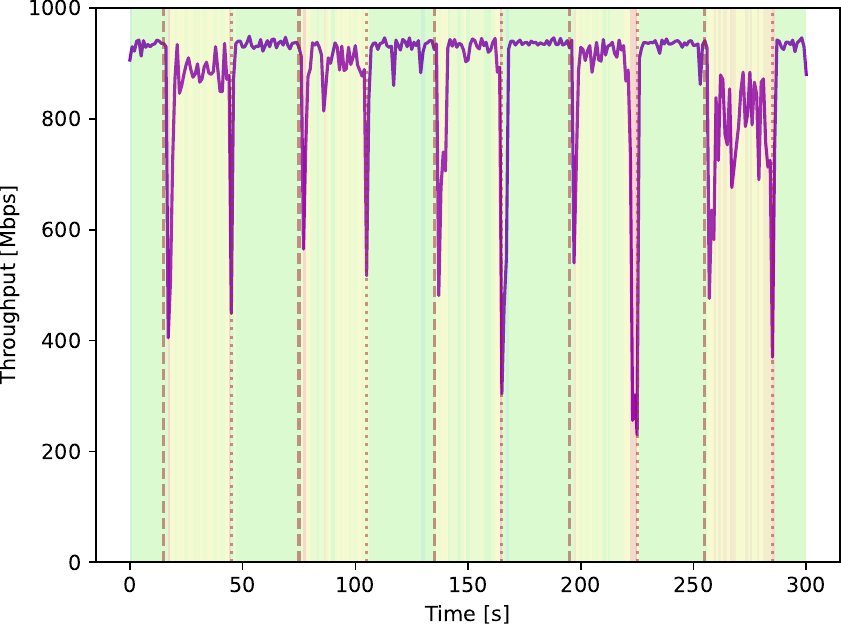}
        \caption{\texttt{iPerf}}
        \end{subfigure}
        \caption{Performance with \gls{HMD}-side moving between \texttt{los} and \texttt{nlos1}.}
        \label{fig:nlos1}
    \end{minipage}
    \hfill
    \begin{minipage}[b]{.325\textwidth}
        \centering
        \begin{subfigure}[t]{\linewidth}
            \includegraphics[width=\linewidth]{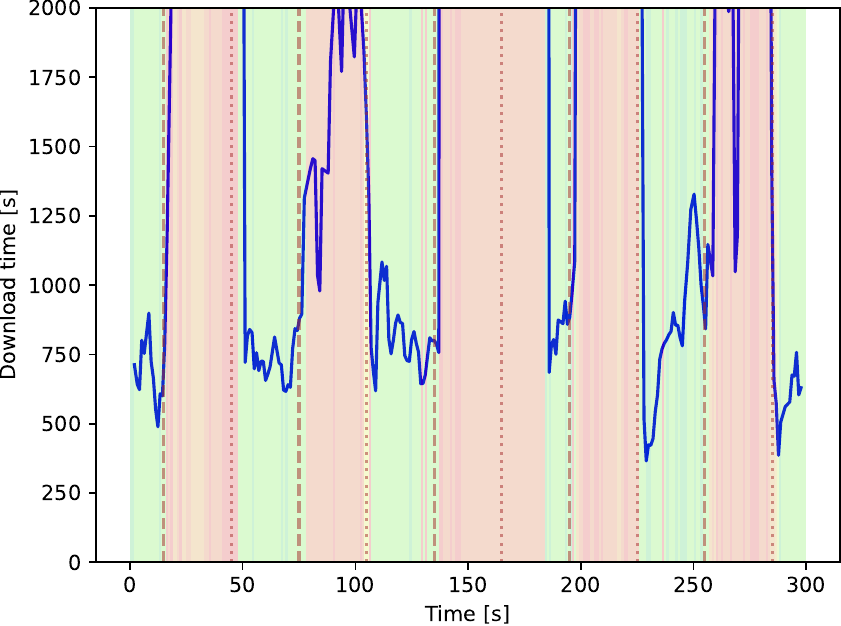}
            \caption{VR streaming}
            \end{subfigure}\\
            \begin{subfigure}[t]{\linewidth}
            \includegraphics[width=\linewidth]{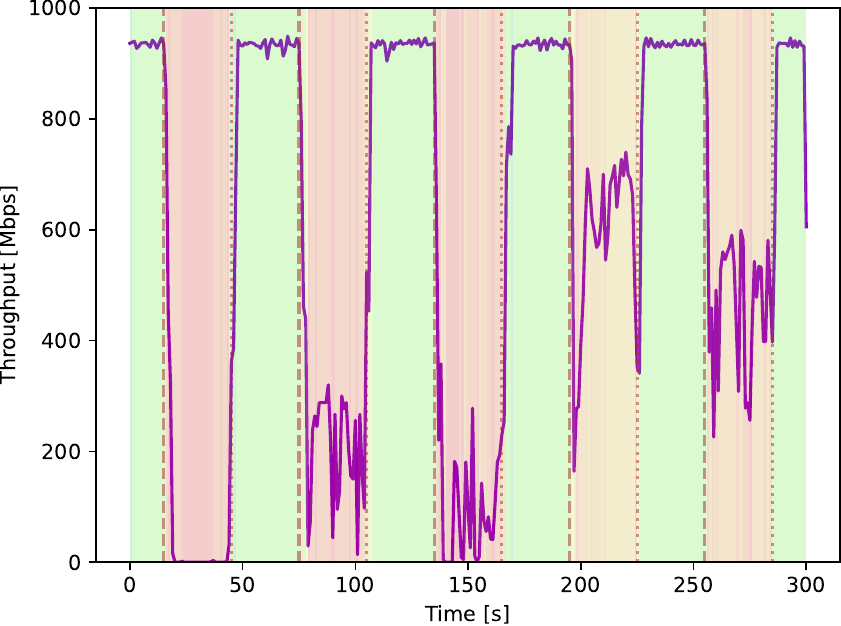}
            \caption{\texttt{iPerf}}
            \end{subfigure}
        \caption{Performance with \gls{HMD}-side moving between \texttt{los} and \texttt{nlos2}.}
        \label{fig:nlos2}
    \end{minipage}
    \hfill
    \begin{minipage}[b]{.325\textwidth}
        \centering
        \begin{subfigure}[t]{\linewidth}
            \includegraphics[width=\linewidth]{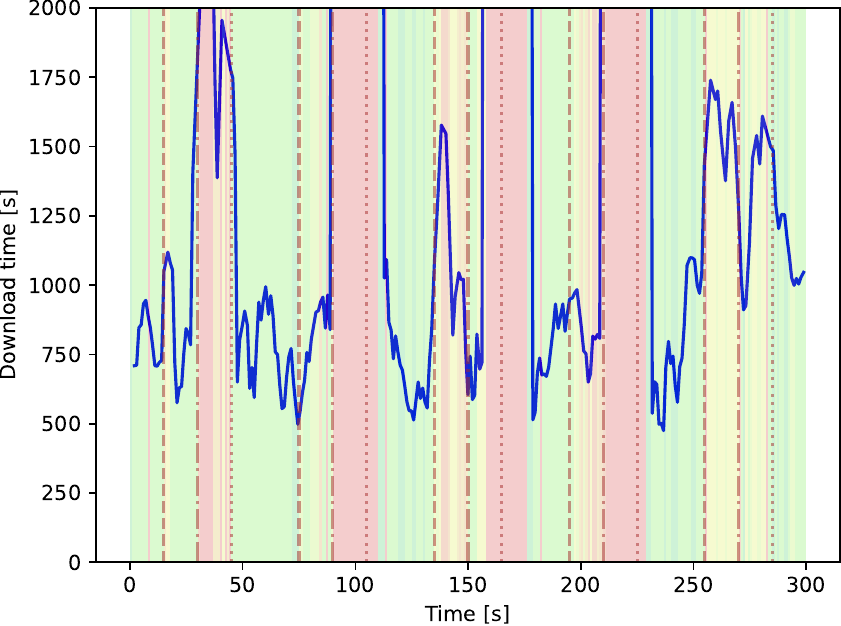}
            \caption{VR streaming}
            \end{subfigure}\\
            \begin{subfigure}[t]{\linewidth}
            \includegraphics[width=\linewidth]{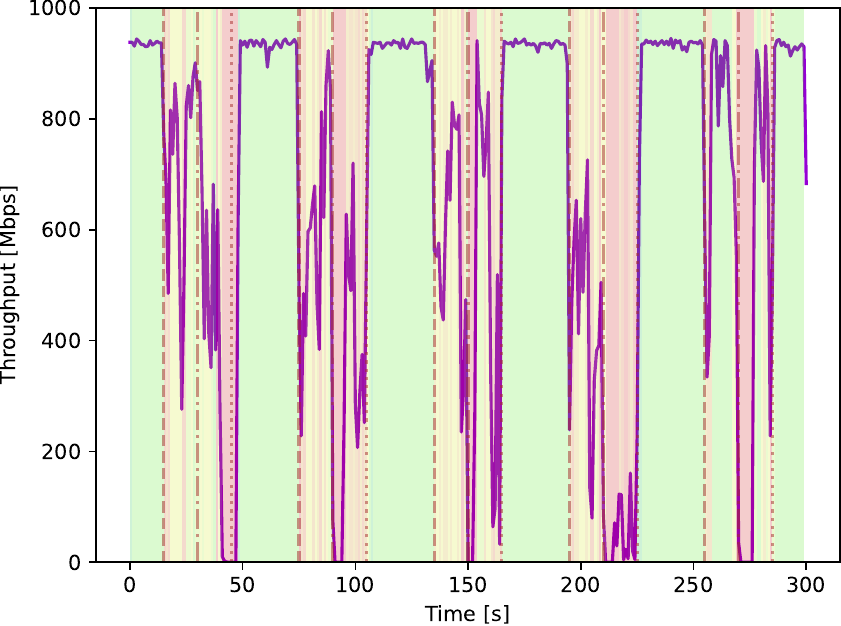}
            \caption{\texttt{iPerf}}
            \end{subfigure}
        \caption{Performance when antenna is intermittently covered with hand.}
        \label{fig:hand}
    \end{minipage}
\end{figure*}
The two NLoS positions were carefully selected such that reception at \texttt{nlos1} was consistently and considerably higher compared to \texttt{nlos2}. As a baseline, we perform static measurements in each of the three positions. As mobility-based experiments, we drive the robot between \texttt{los} and either \texttt{nlos1} or \texttt{nlos2}. Every minute, the robot starts driving away from \texttt{los} at $n'15''$ and starts driving back at $n'45''$. 

As a second type of blockage experiment, we consider self-blockage. An experimenter covers the receive-side antenna with their hand for a part of the experiment, mimicking the situation in which a \gls{VR} user covers their antenna by touching their head. Specifically, starting from $n'15''$, the front and sides of the antenna are covered by holding one's hand in a U-shape. Then, from $n'30''$, all sides are covered by tightly gripping the antenna. Finally, at $n'45''$, the antenna is released.

Finally, to evaluate the differences between mmWave Wi-Fi and legacy Wi-Fi, we repeat these experiments using the routers' legacy Wi-Fi chip. We configure the network to use IEEE 802.11ac, the direct \SI{5}{\giga\hertz} counterpart to IEEE 802.11ad. We use this standard rather than the newer IEEE 802.11ax, as \gls{COTS} devices using its \SI{60}{\giga\hertz} counterpart, IEEE 802.11ay, are not yet available.

To evaluate the performance, we log the throughput for \texttt{iPerf}, and the download time for \gls{VR} streaming. Download time is the inverse of throughput, with \SI{1}{\second} corresponding to, on average, \SI{250}{Mbps}. In addition, we log the transmit \gls{MCS} on the sending router, as this may reveal subtle changes in channel quality not observable from the other metrics.

\section{Evaluation}
In this section, we analyze the results of each experiment, then discuss our overall findings.
\subsection{Experimental Results}
In this evaluation, we evaluate download times for \gls{VR} streaming, where lower is better, and \texttt{iPerf} throughput, where higher is better. In the static baseline, \gls{VR} download times were consistently rather erratic, as we will discuss below. For visual clarity, we therefore smooth the data by taking the average of a 5-sample sliding window. This does not impact any findings. Fig.~\ref{fig:base} shows performance in each of the three static locations. Both tests clearly show that \texttt{nlos1} has only limited impact on performance, while download times degrade severely in \texttt{nlos2}. There are occasional minor drops at \texttt{nlos1}, however performance quickly recovers consistently. In \texttt{nlos2}, throughput is highly erratic, and on average only half of the other two locations. We reiterate that these exact locations were carefully selected to demonstrate both limited and severe degradation due to breaking the \gls{LoS}.

As a secondary baseline, we evaluate performance with legacy Wi-Fi at \SI{5}{\giga\hertz}. As blockage did not impact this performance at all, we only report the results with worst-case mobility and blockage: moving between \texttt{los} and \texttt{nlos2}. Fig.~\ref{fig:legacy} shows the results. Brown lines indicate the start of motion from (dashed) or towards (dotted) \texttt{los}, showing that blockage does not impact performance at all. Signals can penetrate the metal cabinet, and strong reflections help to bypass the blockage entirely.

Following the baselines, we first consider the experiment where the \gls{VR} user travels between \texttt{los} and \texttt{nlos1}. Fig.~\ref{fig:nlos1} does not show a clear difference between the two positions, as is to be expected from the previous results. Inspecting \gls{MCS} measurements does reveal an effect of the blockage however. In the graphs, the background color indicates the current transmit \gls{MCS} of the sending router, with the darkest red being the lowest and the darkest green being the highest. This shows how the \gls{MCS} is consistently decreased when at \texttt{nlos1}. However, as maximum \gls{MCS} corresponds to a PHY-level throughput of roughly \SI{2.3}{Gbps}, even a significant reduction in \gls{MCS} will not impact actual performance, as long as the \SI{1}{Gbps} Ethernet links remain the bottleneck. The \texttt{iPerf} results do reveal a brief decrease in throughput during every motion (in either direction). This is indicative of the motion causing beam misalignment, which is quickly resolved through beamforming. We also note that download times for streaming are consistently more variable compared to the \SI{5}{\giga\hertz} baseline. However, minimum download times, which are mostly smoothed out by the moving window in the plots, were significantly lower with mmWave, reaching as low as \SI{350}{\milli\second}, \SI{100}{\milli\second} less than the absolute minimum download times with legacy Wi-Fi. We hypothesize that the variable download times with mmWave are due to TCP congestion control misbehaving, as is known to occur~\cite{MmWaveTCP1,MmWaveTCP2}.

As expected, performance degradation is considerably more severe when moving between \texttt{los} and \texttt{nlos2}. At \texttt{nlos2}, \gls{MCS} drops towards the lowest possible values, and throughput drops accordingly, as shown in Fig.~\ref{fig:nlos2}. Occasionally, the mmWave link breaks entirely and throughput drops to zero. In one case, the third trip to \texttt{nlos2} with \gls{VR} streaming, it even took 20 seconds to restore the link after returning to \texttt{los}. This clearly illustrates the need for consistently rapid, and even proactive handovers to other routers.

Next to mobility-induced blockage , we also consider self-blockage, such as a \gls{VR} user accidentally covering their head-mounted antenna. Fig.~\ref{fig:hand} shows the impact of self-blockage. An experimenter covered the front and sides of the antenna with their hand starting from each dashed line, and additionally covered the back as well starting from each dash-dotted line. The results show that this blockage degrades performance significantly. Notably, also covering the back-side of the receiving antenna (facing away from the transmitter) degraded performance even further, consistently leading to lower \gls{MCS} and occasionally breaking the link entirely. Furthermore, even very slightly moving one's hand could drastically impact the level of degradation, which partially explains the significant variation in performance between repeats.

\subsection{Discussion}
The above experiments have demonstrated both opportunities and challenges for mmWave wireless connected \gls{VR}. The raw throughput, and therefore minimal stream times mmWave is able to deliver, is vastly superior to lower frequencies, even with gigabit Ethernet interfaces significantly bottlenecking mmWave performance. There are however also a number of challenges, summarized as follows:
\begin{itemize}
    \item \textbf{Throughput dips during motion}: When moving between two high-throughput positions, throughput briefly dips by about \SI{50}{\percent}. This suffices to cause brief freezes in content and should therefore be avoided. This was most likely caused by having to wait for beamforming to trigger and complete. A more rapid, or even proactive beamforming approach, possibly based on \gls{HMD}-side motion detection, may solve this shortcoming~\cite{covrage}.
    \item \textbf{Severe degradation without \gls{LoS}}: In some non-\gls{LoS} positions, throughput was severely degraded, or sometimes even reduced to zero. This clearly has a devastating impact on \gls{QoE}. One solution would be to deploy multiple mmWave routers in different locations with a smart handover system, maximizing the probability of a \gls{LoS} link existing to at least one connected router~\cite{leap}.
    \item \textbf{Jittery streaming performance}: Per-segment download times were highly variable with mmWave, more so than with legacy Wi-Fi. This may be due to TCP congestion control over-correcting for brief reductions in throughput, for example during fluctuating channel conditions or signalling overheads. Careful adjustment of transport and application layer parameters may alleviate this.
\end{itemize}
\section{Conclusions}
In this paper, we experimentally investigated the capabilities and challenges of streaming \gls{VR} over mmWave. We show that, if challenges due to blockage and mobility are addressed, mmWave has the ability to offer vastly higher quality than legacy Wi-Fi, given its potential for extremely high, multi-gigabit throughput. We gathered these findings using a novel testbed for \gls{VR} over mmWave using \gls{COTS} components, where the device representing the \gls{VR} user can be configured to repeatably and accurately perform motion patterns. In future work, we intend to expand on these experiments in several ways, including more extensive motion patterns, and novel streaming frameworks based on UDP, such as QUIC.
\section*{Acknowledgment}
Jakob Struye and Maria Torres Vega are funded by the Research Foundation Flanders (FWO), grant
numbers 1SB0719N and 12W4819N. (Part of) this research was funded by the ICON project INTERACT, realized in collaboration with imec, with project support from VLAIO (Flanders Innovation and Entrepreneurship). Project partners are imec, Rhinox, Pharrowtech, Dekimo and TEO. This research is partially funded by the FWO WaveVR project (Grant number: G034322N). 

\bibliographystyle{IEEEtran}
\bibliography{mmwave_vr_experiments}

\end{document}